\documentclass{llncs}
\usepackage{float}
\usepackage{graphicx}
\usepackage{longtable}
\usepackage[titletoc]{appendix}
\usepackage{pdflscape}

\usepackage{booktabs}
\usepackage[square]{natbib}
\usepackage{pbox}

\usepackage{array}
\newcolumntype{T}[1]{>{\centering\arraybackslash}p{#1}}

\begin{document}

\title{Ecosystem: A Characteristic Of Crowdsourced Environments}

\author{Anamika Chhabra\inst{1} \and S. R. Sudarshan Iyengar\inst{1} \and Poonam Saini\inst{2} \and Rajesh Shreedhar Bhat \inst{3} \and Vijay Kumar \inst{3}}
\institute{{Dept of Computer Science and Engineering, IIT Ropar, Punjab, India }\\
\email{anamika.chhabra@gmail.com}, \email{sudarshan@iitrpr.ac.in}
\and
{Dept of Computer Science and Engineering, PEC University of Technology, India }\\
\email{poonamsaini@pec.ac.in}
\and
{Dept of Information Science and Engineering, PESIT, Bangalore, India}\\
\email{rajeshbhatpesit@gmail.com}, \email{vijay.pesit@gmail.com}
}
\maketitle

\begin{abstract}

The phenomenal success of certain crowdsourced online platforms, such as Wikipedia, is accredited to their ability to tap the crowd's potential to collaboratively build knowledge. While it is well known that the crowd's collective wisdom surpasses the cumulative individual expertise, little is understood on the dynamics of knowledge building in a crowdsourced environment. A proper understanding of the dynamics of knowledge building in a crowdsourced environment would enable one in the better designing of such environments to solicit knowledge from the crowd. Our experiment on crowdsourced systems based on annotations shows that an important reason for the rapid knowledge building in such environments is due to variance in expertise. First, we used as our test bed, a customized Crowdsourced Annotation System (CAS) which provides a group of users the facility to annotate a given document while trying to understand it. Our results showed the presence of different genres of proficiency amongst the users of an annotation system. We observed that the ecosystem in crowdsourced annotation system comprised of mainly four categories of contributors, namely: Probers, Solvers, Articulators and Explorers. We inferred from our experiment that the knowledge garnering mainly happens due to the synergetic interaction across these categories. Further, we conducted an analysis on the dataset of Wikipedia and Stack Overflow and noticed the ecosystem presence in these portals as well. From this  study, we claim that the ecosystem is a universal characteristic of all crowdsourced portals.
\end{abstract}

\textit{``Satellites would someday bring the accumulated knowledge of the world to your fingertips (Arthur C. Clarke, 1970)."}
\section{Introduction}\label{sec:Introduction}
Over the last decade, crowdsourcing has gained immense popularity in the area of learning and collaborative knowledge building, due to the ubiquity of the internet\cite{Estelles-Arolas2012}. The process and the mechanism of knowledge building have now progressed from being a scholarly pursuit of a single person to a decentralized collaborative endeavour of the many.  In the recent past, there has been a considerable amount of effort towards developing appropriate platforms to help the knowledge building and learning process through crowdsourcing\cite{Beldarrain2006}\cite{Bryant2006}. Wikipedia, Quora and StackOverflow are good examples of such platforms. Wikipedia has perhaps evolved to be the best knowledge building experiment mankind has ever attempted in the past two millennia \cite{Korfiatis2006}\cite{Pentzold2006}\cite{Wagner2006}. Withstanding a whole lot of controversies and criticism, this open and free-for-all knowledge database stands tall in terms of its usage and reliability \cite{Giles2005} and has resulted in some of the proprietary encyclopedias go obsolete.\\

Unlike Quora and Stack Overflow which are based on the discussion forum/ Q and A styled approach, Wikipedia is based on what is today called the wiki technology, allowing users to not only have access to its content but also enabling them to add/edit/correct the content online \cite{Raitman2005}. An increasing number of communities benefit from using these crowdsourcing platforms in order to learn new things.\\

While crowdsourcing based platforms are on an a rising scale in rapidly accumulating knowledge\cite{Howe2006a}, the dynamics of knowledge building and learning in such environments is relatively unexplored and not well understood. Understanding the behaviour and the type of interaction of users on knowledge building platforms is an important issue because it affects how we build such platforms in the future. \\

In order to understand the user behaviour on crowdsourced systems, we first developed a customized crowdsourced system Crowdsourced Annotation System (CAS), which combines the features of both annotations and discussion forums. We purposely created this system as we believe that displaying user interaction in the form of annotations on the same page eliminates the problem of navigating from one page to the other in search of the information\cite{Zyto2012}\cite{Thomas2002}. Based on the analysis of the data collected from the experiment on CAS, we present evidence that the reason for rapid knowledge building on such portals is the existence of ecosystem among these categories and that most of the people in the crowd specialize in any one of these categories, and a very small fraction of people behave as multi-specialists. We then present a taxonomy of user behaviour and a model of knowledge building for crowdsourced annotation environments.\\

After having noticed the ecosystem existence in the customized annotation environment, we then conducted an analysis on the data of Wikipedia and Stack Overflow. Wikipedia being based on wiki technology and Stack Overflow being based on a discussion forum based technology provide a different set of activities to its users respectively. It was fortifying to observe the presence of ecosystem on both these portals as well. This observation leads to ecosystem being an inherent quality of successful crowdsourced portals.\\

\section{Related Work}\label{sec:related}

\subsection{Knowledge Building and Learning}
The idea of collaborative knowledge building was first proposed by Scardamalia and Bereiter\cite{scardamalia1994computer}.  Gerry Stahl\cite{Stahl2000} presented a model for collaborative knowledge building which considers the relationship of processes associated with individual minds to those considered to be socio-cultural. Cress and Kimmerle\cite{Cress2008} proposed a theoretical model to describe the process of knowledge building in a wiki environment. Kimmerle, Moskaliuk and Cress\cite{Kimmerle2011} outlined theories related to individual processes of learning and collaborative processes of knowledge building. Nonaka\cite{nonaka1994dynamic} developed a model of knowledge creation which provides an analytical perspective on four patterns of interaction involving tacit and explicit knowledge. Zhang et al.\cite{Zhang2007} conducted experiments on fourth grade students to understand that young students can take collective responsibility for their own knowledge advancement through the use of online tools.\\

Although, there have been several studies conducted to understand what motivates the crowd to contribute in a crowdsourced knowledge building environment \cite{Cress2008}\cite{Nov2007}\cite{forte2008people}, ours is the first study that explores the presence of variance of expertise in the crowd and its importance in catalysing the knowledge building process. 
\subsection{Interfaces for Learning}
Hmelo-Silver, C. E., and Barrows, H. S.\cite{Hmelo-Silver2008} discussed about the need to create appropriate opportunities in order to support student learning and collective knowledge building. Vonderwell and Zachariah\cite{Vonderwell2005} observed that online learner participation and patterns of participation are influenced by various factors such as technology, interface characteristics and student roles. Hence, there is a need to develop pedagogically user-friendly online course interface and management systems. On a similar line, Bederson\cite{Bederson2004} presented various characteristics of flow within the context of interface design with the goal of understanding the type of interfaces that would be most conducive to the users.\\

In online learning environment, the absence of an actual instructor necessitates an interface which could enable the students to understand the complex material in an easy way. Hence, the interface should be designed to maintain the user’s flow and to provide a better learning experience to its end users, thereby, increasing the social inclusion\cite{Bederson2004}. 
\subsection{Annotation Systems}
The technique of discussion forums is the typically adopted means to gather knowledge from the crowd. Participation in discussion forums in online learning websites is linked to higher academic performance\cite{Davies2005}\cite{Deslauriers2011a}. It has also been found that discussions increase the engagement among students, thereby, leading to more learning\cite{Palmer2008}. Nevertheless, discussion forums pose certain limitations too. For example, the typical non-linear branching structure of online discussion forums is insufficient for the realization of truly conversational modes of learning\cite{Thomas2002}.\\

Many researchers have been working on improving the interface so as to achieve maximum benefits out of a crowdsourced platform. In order to eliminate the limitations of discussion forums, Guzdial and Turns\cite{Guzdial2000} defined a specific type of discussion forum which is a computer-mediated anchored discussion forum (CaMILE) with the potential to enable sustained on-topic discussions. The objective has been achieved by embedding hyperlinks from places of discussion into the HTML documents.  Although, it is a considerable step towards improving the detached nature of discussion forums, it still requires the user to follow the link and redirects them to a different context. Cadiz, Gupta and Grudin\cite{Cadiz2000}examined the benefits of in-context annotations. The authors conducted experiments on the members of a large development team using Microsoft Office 2000. \\

Although, some annotation systems are already in place, which emphasize the advantages of in-place annotations, no significant work seems to be done in understanding the dynamics of knowledge building in a crowdsourced annotation environment. The initial work in the area of annotation systems includes work done by Marshall and Brush\cite{Marshall2002}. They designed WebAnn System which utilizes the advantages of in-place annotations by displaying the students’ annotations in the document margins. It allowed readers to see the document and the discussions on the same page. However, due to the fact that people were less comfortable reading online when WebAnn was developed, it was used by them merely as a record keeping tool. People used to read the document offline, mark the annotations and then get back to the system to register them. NB System\cite{Zyto2012} is theoretically similar to the system developed in our work, in the sense that it also provides the facility of in-place annotations. It is a web based tool where users can read and annotate PDF documents using standard web browsers. By the time NB was developed, people had become more technologically advanced and comfortable reading online. It further helped mitigating the obstructions that WebAnn faced. However, there was no discussion or experiments on annotators’ diverse expertise and their behaviour. Su, Yang, Hwang, and Zhang\cite{Su2010} developed a Web 2.0 based collaborative annotation system, PAMS 2.0 (Personalized Annotation Management System 2.0) where the students can create and share individual annotations with annotated documents.
\section{Method}

\subsection{Software}
\begin{figure}
\centering
\includegraphics{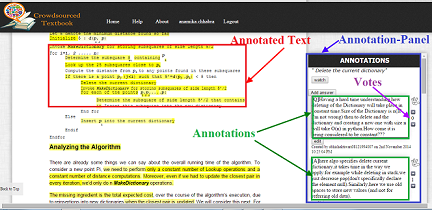}
\caption{A snapshot of CAS (Can be accessed at http://115.248.248.12/CAS/)}
\label{fig:CAS}
\end{figure}

CAS was created using PHP and deployed on the institute server and was accessible over the internet using a web-browser. The portal provided a facility to upload a text document, which the participants could annotate. Any part of the document could be annotated by selecting the text that was to be annotated. A pop-up box then appeared, in which the users were asked to enter their annotations. While adding the annotations, CAS provided the users to select the type of annotation that they were adding, from a drop-down menu. The options that the menu provided were: \textit{[Q] Question, [A] Answer, [I] Insight and [P] Pointer}. [Q] and [A] types of annotations were useful in creating a dialogue between the annotators. The annotators could add any extra information about a part of the content by making use of [I] type annotations. [P] type annotations could be used to point to a similar and useful content on the web. These options had been provided after observing the kind of annotations that users usually add on an annotation system. In order to accommodate any other type of annotation, one more option by the name of ‘Others’ was also provided in the drop-down menu. The annotated text of the main-article was shown highlighted on the interface. When the user clicked on this highlighted text, the corresponding annotations appeared on the right-side panel (see Figure~\ref{fig:CAS}). Also, the system stored all the transaction logs in a back-end database for analysis at any point of time.
 
\subsection{Participants}
Participants in the experiment were 60 second year undergraduate students from the Indian Institute of Technology Ropar, India in the age group of 19 to 21. 41 students were from Computer Science and Engineering, while 19 students were from Electrical Engineering Department. This particular mix of students could help in creating a group of experts and non-experts.

\subsection{Procedure}
The experiment involved annotating an online document, henceforth called the main-article, over a period of 4 days by the participants, through the use of CAS. The participants had to undergo a one-time sign up process before they could start annotating. The main-article that was uploaded was ``Randomized Closest Pair Problem'' which was part of their first course in data structures. The article was part of an advanced chapter from a reference book for this course. The article was not covered in the classroom by the instructor of this course (who is also the second author of this paper) and was chosen carefully for this experiment for three reasons: (1) The article was less straightforward and was mathematically very involved, (2) The authors felt that the students were less likely to understand the uploaded article by independent study, and (3) It was believed that due to no particular prior knowledge on the topic, the group of students are more likely to reflect the data collected from the crowd to a greater extent.
The task for the participants was to understand the main-article through CAS. They were not allowed to talk to each other; they could only see each other’s annotations. Further, they were asked to explicitly mention the type of annotation that they were going to add. They were provided with the option to reply on a [Q] type annotation with an [A] type annotation. They could up vote/down vote an annotation. The annotations in the annotation-panel were displayed in the decreasing order of the votes received, with the annotations with higher votes appearing on the top.

\subsection{Observations}
Following are some of the important observations from our experiment:
\begin{itemize}
\item Over a span of 4 days, 60 students posted 1836 annotations. Out of these, 444 were I-Type annotations, 339 Q-Type, 953 A-Type, 92 P-Type annotations and 8 annotations in the `Others' category. It should be noted that although an option of `Others' was provided; a negligible number of annotations got added in that category. Hence, for our further analysis, we would consider annotations of these four types only. Also, there were 15932 reading entries, 811 voting entries and 66 watchers' entries in the log. A Session is a group of interactions that take place on the website within a given time frame. For example a single session can contain multiple screen or page views, events and social interactions. Also, Pageviews represent the total number of pages that users looked at, on the portal. There were a total of 875 sessions with an average session duration of 15.49 minutes and 3159 pageviews.
\item The first column in the table below indicates the annotation type, the second column is the percentage distribution, third, fourth and fifth columns represent the maximum, minimum and mean number of annotations of a particular type, taken over all the 60 users respectively. The last column indicates the standard deviation from the mean.
\end{itemize}

\section{Analysis of User Contribution}
From Table~\ref{fig:stat_table}, we note that the [A] type annotations were more than half of the total number of annotations, which was 1828, this is due to the fact that a question triggers several back and forth exchange of ideas in the form of [A] type annotations. With just 5\% of the total annotations, [P] type annotations were the least in number amongst the four types. This was still a very significant number, given that they were all links to external articles. With the main-article comprising of 250 lines, 92 [P] type annotations amount to an average of one pointer every 3 lines.
\begin{table}
\centering
\includegraphics{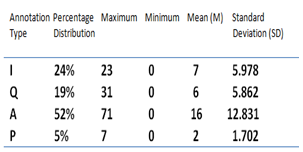}
\caption{Distribution, Max, Min, Mean and SD of Annotation Types}
\label{fig:stat_table}
\end{table}

\begin{figure}
\centering
\includegraphics{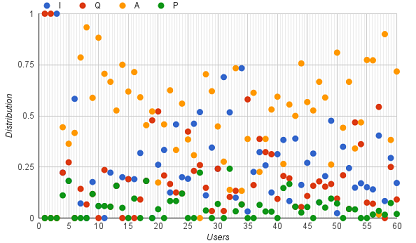}
\caption{Plot of user effort distribution}
\label{fig:Effort}
\end{figure}

After taking note of the [I], [Q], [A] and [P]’s percentage distributions being 24\%, 19\%, 52\% and 5\% respectively, one would infer that most of the users mainly expended their time and effort in answering questions with [A] type annotations and secondarily on posting insights [I] and questions [Q]. This is quite in contrast to what was observed in our logs. The Figure~\ref{fig:Effort} denotes the distribution of [I], [Q], [A] and [P] for individual user contribution of annotations, the x-axis denotes all the 60 users sorted in the increasing order of the number of annotations that they posted. There are 4 dots, colored blue, red, yellow and green, denoting the [I], [Q], [A] and [P] distributions respectively, of individual users. E.g., the 59th user (second last on the X-axis) contributed 68 annotations with [I]=20, [Q]=17, [A]=26 and [P]=5 amounting to the distribution (0.29, 0.25, 0.38, 0.08) which is denoted by 4 dots on the graph with x=59, namely blue: (59,0.29) red: (59,0.25) yellow: (59,0.38) green: (59,0.08). \\

In simple terms, the graph below represents the distribution of efforts by individual users in posting 4 types of annotations. The [A] type annotations across all 60 users has M=16 and SD=13, this is well reflected in the plot below where we observe that the yellow dots are unevenly scattered and are not clustered along any line parallel to X-axis. This is true of red and blue dots as well.

In order to verify the efficacy of CAS, a web-based Feedback Form having 15 questions was prepared. Technology Acceptance Model (TAM) was used to prepare the questions \cite{davis1989perceived}.  On an average, 82.69 percent students gave `Strongly Agree' or `Agree' as their answers. The users also reported that the introduction of various types of annotations (i.e. I, Q, A, P) in CAS was useful in sharing the knowledge in the group. Also, displaying the annotations on the same page as the text was useful in bringing the students' attention towards annotations added by others, which further enhanced their knowledge. In \cite{chhabra2015framework}, the authors discuss the efficacy of CAS and its usefulness in an educational environment. 

\section{The Presence of an Ecosystem in CAS}
We call a user k-unispecialist if s/he appears in the list of top k contributors (in terms of number of annotations) for precisely one type of annotation, and does not appear in the top k of the other three types. E.g., a user A with the annotations [I]=7, [Q]=22, [A]=16, [P]=2 is ranked 26th, 2nd, 23rd and 14th in [I], [Q], [A] and [P] respectively. Here, A is 2-unispecialist as s/he appears as one of the top 2 contributors in precisely one type and doesn’t appear in the other 3 types. We note that, by definition, A is a k-unispecialist with k=3, 4... 13. We similarly define a k-bispecialist who appears in the list of top k contributors in precisely 2 types of annotations. E.g., the user A in the above example is a 14-bispecialist but not a 23-bispecialist. On the same lines, we define a k-Trispecialist and a k-Quadspecialist. \\

\begin{figure}
\centering
\includegraphics{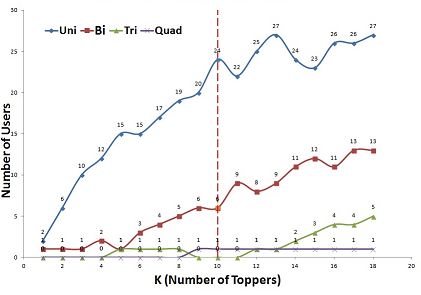}
\caption{Existence of Uni/Bi/Tri/Quad Specialists}
\label{fig:unibitriquad}
\end{figure}

The Figure~\ref{fig:unibitriquad} shows a plot of the number of k-uni/bi/tri/quad specialists with x-axis running through k = 1, 2, 3 ... 18. At k=10 (along the red dotted line) we observe that the number of 10-unispecialists, 10-bispecialists, 10-trispecialists and 10-quadspecialists are 24, 6, 0 and 1 respectively. The plot indicates that the top contributors are proficient in posting only a particular type of annotation. 
 
There are several unispecialists (blue line in the plot) and very few bispecialists (red line) and negligibly few trispecialists (green) and quadspecialists (purple). It is this ecosystem that exists in a crowdsourced environment that fosters knowledge building and guarantees both - quantity and quality of information. There are Explorers who are good at pointing to external resources which helps garner more data for the users; there are Solvers who are good at answering questions and Probers who ask questions which instigates the crowd to think outside the realms of the given article. Articulators with their above average ability for expressive writing, play a good role in paraphrasing parts of the document which are perceivably less clear to the readers.\\

\begin{figure}
\centering
\includegraphics{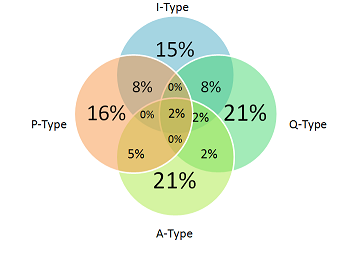}
\caption{Percentage distribution of  I, Q, A and P-Type Annotations at K=13}
\label{fig:Percentage}
\end{figure}

Figure~\ref{fig:Percentage} shows the percentage distribution of the four types of categories at a certain value of K (K=13 here). This value of K has been randomly chosen. The other values of K (up to a threshold point) exhibit the same behaviour. We observe that when 13 top performers from all the four categories are taken together, then out of these, 73\% of the users show expertise only in a particular category, i.e. they behave as uni-specialists. Moreover, only 2\% of all the users post annotations in all the four categories. The figure also shows the percentage distribution in other combination of categories. For example, only 8\% of all the users added both I-Type and Q-type annotations. Only 8\% added both I-Type and P-Type annotations. Only 5\% added both A-Type and P-type annotations. And only 2\% added both Q-type and A-Type annotations. \\

One can also observe that none of the annotators added in I, Q and P taken together, I, P and A taken together, and P, A and Q taken together respectively. Further, Only 2\% added in I, Q and A categories taken together.

\section{A Model of User Behaviour in Annotation Systems}
Based on the observations from the experiment, we can divide the annotators into various categories. Following are the main types of activities that annotators exhibit while collaborating on the portal:
\begin{enumerate}
\item \textit{Articulators}: The users who keep adding their insights about the particular text they are reading, and help in better understanding of the text, fall into this category. The annotations added by them are called I-Type Annotations.
\item \textit{Probers}: The users who ask a lot of questions are called Probers. They might indicate a class of users who have less knowledge of the given text as well as people, who are in general inquisitive, and have this knack of asking good questions. The annotations added by them are called Q-Type Annotations.
\item \textit{Solvers}: The users who answer the questions posted by others are called Solvers. As compared to the Articulators, who keep adding their knowledge even without having being asked any question, this type of users get into action only when they see a question. The annotations added by them are called A-Type Annotations.
\item \textit{Explorers}: The users who do not restrict themselves only to the current resource, but also keep looking for some relevant information outside the given text, and make it available to everyone, go to Explorers category. The annotations added by them are called P-Type Annotations, where P indicates pointers.\\
We also separately define one more type of users:
\item \textit{Voters}: The users who do not add any annotations, just read and up vote or down vote the annotations added by others. \\
Finally, for the sake of completeness, we describe yet another category of users, who are passive lurkers and do not take any of the above mentioned roles.
\item \textit{Viewers}: The users who only read the annotations added by others. Although these users make use of the portal in their own learning, they do not contribute in the knowledge building process on the portal. 
\end{enumerate}

\begin{figure}
\centering
\includegraphics{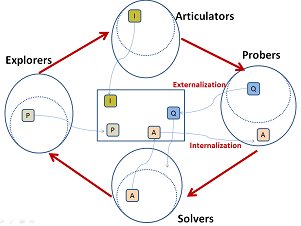}
\caption{The rectangle represents the Annotation System. The Ovals represent the Cognitive Systems of the annotators. The Prober has been shown externalizing Q-type annotation to the system and internalizing A-Type annotation}
\label{fig:model}
\end{figure}

We now present a theoretical model of collaborative knowledge building in annotation systems by assuming a systemic perspective\cite{Cress2008}. According to Luhmann, for the knowledge building and learning to take place, communication between the social system (the annotation system here) and the cognitive system (the annotators here) happens\cite{luhmann1995social}. We observe that the cognitive systems of the annotators externalize  or internalize  in order to learn or add to the knowledge building respectively. Figure~\ref{fig:model} illustrates the processes of Externalization and Internalization happening among the categories. Articulators, Probers, Solvers and Explorers post I, Q, A and P-type annotations respectively, through the process of Externalization. For example, the cognitive system of a Prober asks a question by externalising his question in the form of a Q-type of annotation to the system. This Q-type of annotation is perceived by the cognitive system of the Solver through the process of Internalization. And, in response, the solver externalizes the answer in the form of an A-Type of annotation, which is then internalized by the Prober.\\

The figure also shows a chain of Ecosystem that exists among the categories. For example, the I-type of annotation of an Articulator prompts a Prober to ask a question through Q-type of annotation, which in turn provokes a Solver to externalize through A-type of annotation, seeing which the Explorer looks to the outside resources and adds a P-Type of annotation.

\section{Ecosystem Existence in Wikipedia}
After having witnessed the existence of ecosystem in crowdsourced annotation environments, we conducted an analysis on the data of Wikipedia.  The types of activities that the Wikipedia contributors perform are different from those performed by the users of an annotation system. We wanted to observe whether the users of Wikipedia performed a mixture of all these activities, or they specialised in any one of them only. In order to check this, we analysed the edit history of ten Wikipedia articles from diverse areas. The articles observed were: Alan Turing (AT), Albert Einstein (AE), Barack Obama (BO), Fermat's Last Theorem (FLT), India (I), Leonhard Euler (LE), Mathematics (M), Sachin Tendulkar (ST), Srinivasa Ramanujan (SR) and Stirling Number (SN). We examined the comments associated with top 5000 edits for each article and observed that the main types of activities performed by the contributors of Wikipedia are: adding new text, making changes in the previously added text, updating the existing content with the latest information, reverting to the previous state (for example in case of vandalism) and providing references for the existing content. Based on this observation, we divided the contributors into: \textit{Editors} (E), \textit{Reverters} (R), \textit{Adders} (A), \textit{Updaters} (U) and \textit{Pointers} (P) respectively. The discussions taking place in the `talk pages' were not considered while forming the categories in this study, considering them to be the background actions happening due to the activities already considered. Now, the goal was to identify the percentage of contributors that lie in more than one of these categories.

\begin{figure}[htbp]
\centering
\includegraphics{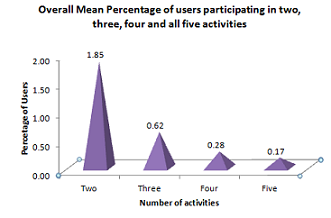}
\caption{Overall Mean Percentage of users participating in two, three, four and all five activities}
\label{fig:wiki_eco}
\end{figure}

Tables \ref{category_one}, \ref{category_two}, \ref{category_three}, \ref{category_four} and \ref{category_five} in the appendix show the data collected on all ten pages in terms of number of contributors falling in each category as well as in more than one category. It was observed that the percentage of contributors who performed two activities was 1.85\%, those who performed three activities was 0.62\%, those who performed four of these activities was 0.28\% and the contributors performing all five of these activities were only 0.17\%. (See Figure~\ref{fig:wiki_eco})

\begin{figure}[htbp]
\centering
\includegraphics[scale=.9]{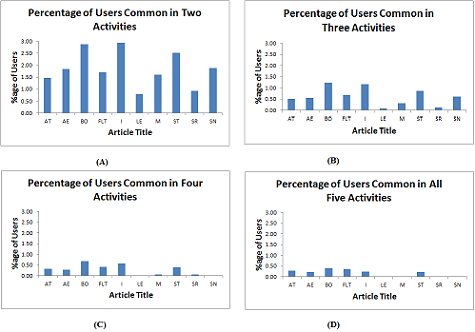}
\caption{Percentage of users who perform more than one type of activity for the ten articles}
\label{fig:wiki_common}
\end{figure}
Figure~\ref{fig:wiki_common} shows the percentage of users who perform more than one type of activity for all the articles respectively.  When two activities are put together, there are ten possible combinations viz. (E, R), (E, A), (E, U), (E, P), (R, A), (R, U), (R, P), (A, U), (A, P) and (U, P). When three activities are put together, there are ten possible combinations viz. (E, R, A), (E, R, U), (E, R, P), (E, A, U), (E, A, P), (E, U, P), (R, A, U), (R, A, P), (R, U, P) and (A, U, P). When four are activities put together, there are six possible combinations viz. (E, R, A, U), (E, R, A, P), (E, R, U, P), (E, A, U, P) and (R, A, U, P). Therefore, parts (A), (B) and (C) of Figure~\ref{fig:wiki_common} show the mean percentages, whereas the part (D) shows the actual percentage of the only combination (E, R, A, U, P) of all five activities put together.

We clearly see that the number of people who perform more than one type of activity is very less and the contributors specialize in one of the activities. This is significant evidence that the ecosystem exists in Wikipedia. We believe that due to this ecosystem, one type of contributors trigger the other type and this recursive process helps in giving a better shape to the Wikipedia articles. 

\section{Existence of Ecosystem in Stack Overflow}
Stack Overflow is an example of a crowdsourced knowledge building portal based on discussion forums. It is a question and answer site for programmers. We analysed the main kinds of activities that the users perform on the site. It was observed that the users mainly ask questions, answer the questions asked by others, read and vote others’ questions or answers or edit the questions posted by others. Based on this observation, we divided the users on the site into the four categories: 1) Questioners 2) Answerers 3) Editors 4) Voters

\begin{figure}[htbp]
\centering
\includegraphics{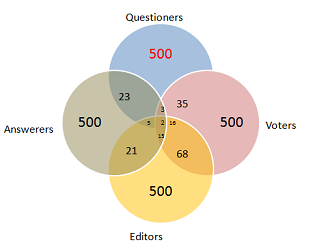}
\caption{Number of users in each category in StackOverflow}
\label{fig:so_venn}
\end{figure}

In order to check whether the users participate more in one of the activities as compared to the other ones, we analyzed the top 500 performers in each category. Then we found the number of common users in these categories, two taken together, three taken together and all four taken together. It was observed that, very few people participated in activities in more than one category. Table~\ref{fig:SO_table} shows the number of users in various categories put together. Figure~\ref{fig:so_venn} shows the Venn diagram of the number of users in each category. It can be seen that only two users participated in all the four activities.  

\begin{table}[htbp]
\centering
\includegraphics{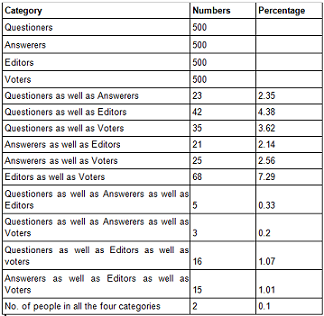}
\caption{Analysis of Stack overflow categories}
\label{fig:SO_table}
\end{table}

\begin{figure}[ht!]
\centering
\includegraphics{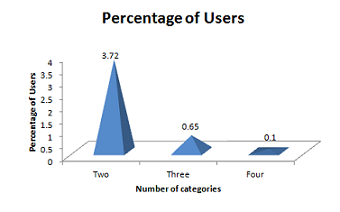}
\caption{Number of users in each category in StackOverflow}
\label{fig:so_percentage}
\end{figure}

Figure~\ref{fig:so_percentage} shows that on an average 3.72\% people perform in two of the categories, 0.65\% people perform in three of the categories, and only 0.1\% people perform in all the four categories.

All these observations give an indication that the presence of ecosystem is a characteristic of successful crowdsourced platforms. We believe that it is certainly one of the important reasons for the success or failure of a portal. Due to the presence of this ecosystem only, knowledge building portals such as Wikipedia, Stackoverflow and Quora have been flourishing and expanding.
\section{Conclusion and Future Work}
In this work, we observed the presence of various categories of participants in the portals based on crowdsourcing. We first developed and executed an experiment involving text annotations by students in a web based environment. We observed that the top contributors specialised in a single task, and further noted that the knowledge building process in a collaborative environment is triggered and fostered by the rapid back and forth exchange of information between the top contributors, who are - as we term - the unispecialists. We report on an empirical evidence for the presence of an ecosystem of expertise in a crowdsourced annotation environment. Our experiment shows that the annotation ecosystem comprises of four types of people: Articulators, Probers, Solvers and Explorers who are good at providing insights, asking questions, posting answers and pointing to external resources respectively. While it is commonsensical to observe ``more the merrier'' holding good in a crowdsourced environment, it takes a detailed experimental investigation to establish the fact that the very reason why ``Whole is greater than parts'' is due to the distribution of expertise in the crowd.\\

Getting inspired from the results of the experiment on annotation system, we then conducted an analysis on the data sets of Wikipedia and Stack Overflow. It was interesting to observe the presence of ecosystem on these portals as well. The results from these analyses give an indication of ecosystem being an inherent characteristic of successful crowdsourced environments. We further believe that an interesting dimension of research would be to observe that an inappropriate mix of participants' expertise might lead to the failure of a portal seeking to gather the services of the crowd.\\

The study has many implications for the portal designers in improving the capability of a portal trying to make use of the power of crowd. We envision that the collaborative knowledge building environments in the future will be designed taking note of the presence of `diversity of expertise’ which is a great catalyst for knowledge building. It is important to identify the categories of contributors and aid them to contribute better. The portal designers may employ strategies that identify the expertise of a particular user and encourage them accordingly, e.g., enabling - at the interface level - a better display of questions to the ``Solvers'' category would help in converting their tacit knowledge into explicit knowledge, thereby, catalyzing the knowledge building process. Further, knowing that the presence of every genre of users is essential in improving the knowledge building process, the portal designers may employ various incentivizing mechanisms making use of measures like votes and badges to further improve the engagement level of different types of users. This will help the ecosystem to flourish even more, thereby helping to tap the full potential of the participants.\\

In the near future, we plan to conduct a longitudinal investigation on CAS for a few months. This may prevent the biases drawn from a limited time span of the experiment, if any. Based on the statistics thus obtained, more qualitative investigation can be performed, the interaction among the annotators can be better analyzed and the categories can be reformed, if required. We also plan to conduct an analysis of the portals which have been unsuccessful in utilizing the potential of the crowd. It would be interesting to further investigate on the exact distribution of the expertise in large crowdsourced environments.

\bibliographystyle{splncs}
\bibliography{ecosystem_springer_bibfile}

\begin{thebibliography}{10}

\bibitem{Estelles-Arolas2012}
Estelles-Arolas, E., Gonzalez-Ladron-de Guevara, F.:
\newblock {Towards an integrated crowdsourcing definition}.
\newblock Journal of Information Science \textbf{38} (2012)  189--200

\bibitem{Beldarrain2006}
Beldarrain, Y.:
\newblock {Distance Education Trends: Integrating new technologies to foster
  student interaction and collaboration}.
\newblock Distance Education \textbf{27} (2006)  139--153

\bibitem{Bryant2006}
Bryant, T.:
\newblock {Social software in academia}.
\newblock Educause quarterly (2006)  61--64

\bibitem{Korfiatis2006}
Korfiatis, N.T., Poulos, M., Bokos, G.:
\newblock {Evaluating authoritative sources using social networks: an insight
  from Wikipedia}.
\newblock Online Information Review \textbf{30} (2006)  252--262

\bibitem{Pentzold2006}
Pentzold, C., Seidenglanz, S.:
\newblock {Foucault@ Wiki: first steps towards a conceptual framework for the
  analysis of Wiki discourses}.
\newblock \ldots of the 2006 international symposium on \ldots (2006)  59--68

\bibitem{Wagner2006}
Wagner, C., Kong, H.:
\newblock {Breaking the Knowledge Acquisition Bottleneck Through
  Conversational}.
\newblock Information Resources Management Journal (IRMJ) \textbf{19} (2006)
  70--83

\bibitem{Giles2005}
Giles, J.:
\newblock {Internet encyclopaedias go head to head.}
\newblock Nature \textbf{438} (2005)  900--901

\bibitem{Raitman2005}
Raitman, R., Augar, N.:
\newblock {Employing Wikis for Online Collaboration in the E-Learning
  Environment: Case Study}.
\newblock Third International Conference on Information Technology and
  Applications (ICITA'05) \textbf{2} (2005)  142--146

\bibitem{Howe2006a}
Howe, J.:
\newblock {The Rise of Crowdsourcing}.
\newblock North \textbf{14} (2006)  1--5

\bibitem{Zyto2012}
Zyto, S., Karger, D., Ackerman, M., Mahajan, S.:
\newblock {Successful classroom deployment of a social document annotation
  system}.
\newblock Proceedings of the SIGCHI \ldots (2012)  1883--1892

\bibitem{Thomas2002}
Thomas, M.:
\newblock {Learning within incoherent structures: the space of online
  discussion forums}.
\newblock Journal of Computer Assisted Learning \textbf{18} (2002)  351--366

\bibitem{scardamalia1994computer}
Scardamalia, M., Bereiter, C.:
\newblock Computer support for knowledge-building communities.
\newblock The journal of the learning sciences \textbf{3} (1994)  265--283

\bibitem{Stahl2000}
Stahl, G.:
\newblock {A model of collaborative knowledge-building}.
\newblock Fourth international conference of the learning \ldots (2000)  70--77

\bibitem{Cress2008}
Cress, U., Kimmerle, J.:
\newblock {A systemic and cognitive view on collaborative knowledge building
  with wikis}.
\newblock International Journal of Computer-Supported Collaborative Learning
  \textbf{3} (2008)  105--122

\bibitem{Kimmerle2011}
Kimmerle, J., Moskaliuk, J., Cress, U.:
\newblock {Using Wikis for Learning and Knowledge Building: Results of an
  Experimental Study.}
\newblock Educational Technology \& \ldots \textbf{14} (2011)  138--148

\bibitem{nonaka1994dynamic}
Nonaka, I.:
\newblock A dynamic theory of organizational knowledge creation.
\newblock Organization science \textbf{5} (1994)  14--37

\bibitem{Zhang2007}
Zhang, J., Scardamalia, M., Lamon, M., Messina, R., Reeve, R.:
\newblock {Socio-cognitive dynamics of knowledge building in the work of 9- and
  10-year-old}.
\newblock Educational Technology Research and Development, \textbf{2} (2007)
  117--145

\bibitem{Nov2007}
Nov, O.:
\newblock {What motivates wikipedians?}
\newblock Communications of the ACM \textbf{50} (2007)  60--64

\bibitem{forte2008people}
Forte, A., Bruckman, A.:
\newblock {Why do people write for Wikipedia? Incentives to contribute to
  open--content publishing}.
\newblock In: Proceedings of 41st Annual Hawaii International Conference on
  System Sciences (HICSS). (2008)  1--11

\bibitem{Hmelo-Silver2008}
Hmelo-Silver, C.E., Barrows, H.S.:
\newblock {Facilitating Collaborative Knowledge Building}.
\newblock Cognition and Instruction \textbf{26} (2008)  48--94

\bibitem{Vonderwell2005}
Vonderwell, S., Zachariah, S.:
\newblock {Factors that influence participation in online learning}.
\newblock Journal of Research on Technology in \ldots \textbf{5191} (2005)
  213--230

\bibitem{Bederson2004}
Bederson, B.B.:
\newblock {Interfaces for staying in the flow}.
\newblock Ubiquity \textbf{2004} (2004) ~1

\bibitem{Davies2005}
Davies, J., Graff, M.:
\newblock {Performance in e�learning: online participation and student
  grades}.
\newblock British Journal of Educational Technology \textbf{36} (2005)

\bibitem{Deslauriers2011a}
Deslauriers, L., Schelew, E., Wieman, C.:
\newblock {Improved learning in a large-enrollment physics class.}
\newblock Science (New York, N.Y.) \textbf{332} (2011)  862--4

\bibitem{Palmer2008}
Palmer, S., Holt, D., Bray, S.:
\newblock {Does the discussion help? The impact of a formally assessed online
  discussion on final student results}.
\newblock British Journal of Educational Technology \textbf{39} (2008)
  847--858

\bibitem{Guzdial2000}
Guzdial, M., Turns, J.:
\newblock {Effective discussion through a computer-mediated anchored forum}.
\newblock The journal of the learning sciences \textbf{9} (2000)  437--469

\bibitem{Cadiz2000}
Cadiz, J.J., Gupta, A., Grudin, J.:
\newblock {Using Web Annotations for Asynchronous Collaboration Around
  Documents Using Web Annotations for Asynchronous Collaboration Around
  Documents}.
\newblock In Proceedings of the 2000 ACM conference on Computer supported
  cooperative work (2000)  309--318

\bibitem{Marshall2002}
Marshall, C.C., Brush, A.J.B.:
\newblock {From personal to shared annotations}.
\newblock CHI '02 extended abstracts on Human factors in computing systems
  (2002)  812--813

\bibitem{Su2010}
Su, A.Y., Yang, S.J., Hwang, W.Y., Zhang, J.:
\newblock {A Web 2.0-based collaborative annotation system for enhancing
  knowledge sharing in collaborative learning environments}.
\newblock Computers \& Education \textbf{55} (2010)  752--766

\bibitem{davis1989perceived}
Davis, F.D.:
\newblock Perceived usefulness, perceived ease of use, and user acceptance of
  information technology.
\newblock MIS quarterly (1989)  319--340

\bibitem{chhabra2015framework}
Chhabra, A., Iyengar, S., Saini, P., Bhat, R.S.:
\newblock A framework for textbook enhancement and learning using crowd-sourced
  annotations.
\newblock arXiv preprint arXiv:1503.06009 (2015)

\bibitem{luhmann1995social}
Luhmann, N.:
\newblock Social systems.
\newblock Stanford University Press (1995)

\end{thebibliography}
\newpage
%\section{\appendixname}
\appendix
\section{APPENDIX}
Tables \ref{category_one}, \ref{category_two}, \ref{category_three}, \ref{category_four} and \ref{category_five} show the data collected on all ten pages in terms of number of contributors falling in one, two, three, four and five categories respectively. Please note that in the tables, short forms of Alan Turing (AT), Albert Einstein (AE), Barack Obama (BO), Fermat's Last Theorem (FLT), India (I), Leonhard Euler (LE), Mathematics (M), Sachin Tendulkar (ST), Srinivasa Ramanujan (SR), Stirling Number (SN) have been used.

\begin{table}[]
\centering
\begin{tabular}{T{2.5cm} T{0.8cm} T{0.8cm} T{0.8cm} T{0.8cm} T{0.8cm} T{0.8cm} T{0.8cm} T{0.8cm} T{0.8cm} T{0.8cm}}
{\bf Category/ Article} & {\bf AT} &  {\bf AE} &{\bf BO} & {\bf FLT} &  {\bf I} &  {\bf LE} &  {\bf M} &  {\bf ST} &  {\bf SR} &  {\bf SN} \\ \hline
{\bf Editors}          & 162               & 267                   & 250                & 90                          & 195         & 75                   & 143               & 130                    & 80                        & 9     \\
{\bf Reverters}        & 340               & 482                   & 437                & 128                         & 374         & 303                  & 429               & 155                    & 213                       & 10                    \\
{\bf Adders}           & 105               & 126                   & 161                & 37                          & 135         & 44                   & 54                & 78                     & 45                        & 7                     \\
{\bf Updaters}         & 19                & 14                    & 47                 & 5                           & 38          & 4                    & 4                 & 47                     & 1                         & 1                     \\
{\bf Pointers}         & 59                & 74                    & 97                 & 29                          & 59          & 24                   & 52                & 41                     & 29                        & 5            
\end{tabular}
\caption{Number of Contributors Falling in One Category}
\label{category_one}
\end{table}
%\end{landscape}

\begin{table}[]
\centering
\begin{tabular}{T{3.5cm} T{0.7cm} T{0.7cm} T{0.7cm} T{0.7cm} T{0.7cm} T{0.7cm} T{0.7cm} T{0.7cm} T{0.7cm} T{0.7cm}}
{\bf Category/ Article} & {\bf AT} &  {\bf AE} &{\bf BO} & {\bf FLT} &  {\bf I} &  {\bf LE} &  {\bf M} &  {\bf ST} &  {\bf SR} &  {\bf SN} \\ \hline

{\bf Editors, Reverters}                             & 28  & 55  & 71  & 17  & 67  & 11  & 46  & 29  & 11  & 0  \\ 
{\bf Editors, Adders}                                & 15  & 40  & 40  & 6   & 35  & 5   & 15  & 17  & 9   & 1  \\ 
{\bf Editors, Updaters}                              & 4   & 5   & 18  & 1   & 12  & 0   & 1   & 6   & 0   & 0  \\ 
{\bf Editors, Pointers}                             & 12  & 19  & 32  & 6   & 18  & 5   & 13  & 9   & 2   & 1  \\ 
{\bf Reverters, Adders}                              & 16  & 26  & 32  & 4   & 37  & 9   & 11  & 17  & 5   & 0  \\ 
{\bf Reverters, Updaters}                            & 5   & 4   & 15  & 2   & 17  & 1   & 1   & 13  & 0   & 1  \\ 
{\bf Reverters, Pointers}                            & 6   & 13  & 29  & 7   & 21  & 3   & 14  & 7   & 4   & 1  \\ 
{\bf Adders, Updaters}                               & 4   & 3   & 14  & 1   & 13  & 0   & 0   & 8   & 0   & 0  \\ 
{\bf Adders, Pointers}                               & 7   & 11  & 23  & 4   & 13  & 1   & 7   & 5   & 3   & 1  \\ 
{\bf Updaters, Pointers}                             & 3   & 2   & 12  & 1   & 3   & 0   & 1   & 3   & 0   & 1  
  
\end{tabular}
\caption{Number of Contributors falling in Two categories}
\label{category_two}
\end{table}
%\end{landscape}

\begin{table}[]
\centering
\begin{tabular}{T{4.9cm} T{0.6cm} T{0.6cm} T{0.6cm} T{0.6cm} T{0.6cm} T{0.6cm} T{0.6cm} T{0.6cm} T{0.6cm} T{0.6cm}}
{\bf Category/ Article} & {\bf AT} &  {\bf AE} &{\bf BO} & {\bf FLT} &  {\bf I} &  {\bf LE} &  {\bf M} &  {\bf ST} &  {\bf SR} &  {\bf SN} \\ \hline
{\bf Editors, Reverters,  Adders}                    & 8   & 19  & 21  & 4   & 23  & 2   & 6   & 10  & 1   & 0  \\ 
{\bf Editors, Reverters, Updaters}                   & 3   & 3   & 11  & 1   & 11  & 0   & 0   & 6   & 0   & 0  \\ 
{\bf Editors, Reverters, Pointers}                   & 5   & 7   & 19  & 5   & 13  & 1   & 5   & 6   & 1   & 0  \\ 
{\bf Editors, Adders, Updaters}                      & 2   & 3   & 10  & 1   & 8   & 0   & 0   & 2   & 0   & 0  \\ 
{\bf Editors, Adders, Pointers}                      & 4   & 7   & 17  & 3   & 11  & 0   & 5   & 3   & 1   & 1  \\ 
{\bf Editors, Updaters, Pointers}                    & 2   & 2   & 9   & 1   & 3   & 0   & 0   & 2   & 0   & 0  \\ 
{\bf Reverters, Adders, Updaters}                   & 2   & 3   & 9   & 1   & 11  & 0   & 0   & 3   & 0   & 0  \\ 
{\bf Reverters, Adders, Pointers}                    & 3   & 5   & 12  & 2   & 10  & 1   & 4   & 3   & 2   & 0  \\ 
{\bf Reverters, Updaters, Pointers}                  & 2   & 2   & 6   & 1   & 2   & 0   & 1   & 2   & 0   & 1  \\ 
{\bf Adders, Updaters, Pointers}                     & 3   & 2   & 7   & 1   & 2   & 0   & 0   & 2   & 0   & 0  
  
\end{tabular}
\caption{Number of Contributors Falling in Three Categories}
\label{category_three}
\end{table}
%\end{landscape}

\begin{table}[]
\centering
\begin{tabular}{T{6.4cm} T{0.6cm} T{0.6cm} T{0.6cm} T{0.6cm} T{0.6cm} T{0.6cm} T{0.6cm} T{0.6cm} T{0.6cm} T{0.6cm}}
{\bf Category/ Article} & {\bf AT} &  {\bf AE} &{\bf BO} & {\bf FLT} &  {\bf I} &  {\bf LE} &  {\bf M} &  {\bf ST} &  {\bf SR} &  {\bf SN} \\ \hline
{\bf Editors, Reverters, Adders, Updaters}           & 2   & 3   & 8   & 1   & 8   & 0   & 0   & 2   & 0   & 0   \\
{\bf Editors, Reverters, Adders, Pointers}           & 3   & 4   & 10  & 2   & 9   & 0   & 2   & 3   & 1   & 0  \\
{\bf Editors, Reverters, Updaters, Pointers}         & 2   & 2   & 6   & 1   & 2   & 0   & 0   & 2   & 0   & 0  \\ 
{\bf Editors, Adders, Updaters, Pointers}            & 2   & 2   & 6   & 1   & 2   & 0   & 0   & 1   & 0   & 0  \\ 
{\bf Reverters, Adders,  Updaters, Pointers}         & 2   & 2   & 4   & 1   & 2   & 0   & 0   & 1   & 0   & 0  

\end{tabular}
\caption{Number of Contributors in Four Categories}
\label{category_four}
\end{table}
%\end{landscape}

\begin{table}[]
\centering
\begin{tabular}{T{7.7cm} T{0.6cm} T{0.6cm} T{0.6cm} T{0.6cm} T{0.6cm} T{0.6cm} T{0.6cm} T{0.6cm} T{0.6cm} T{0.6cm}}
{\bf Category/ Article} & {\bf AT} &  {\bf AE} &{\bf BO} & {\bf FLT} &  {\bf I} &  {\bf LE} &  {\bf M} &  {\bf ST} &  {\bf SR} &  {\bf SN} \\ \hline
{\bf Editors, Reverters, Adders, Updaters, Pointers} & 2   & 2   & 4   & 1   & 2   & 0   & 0   & 1   & 0   & 0 

\end{tabular}
\caption{Number of Contributors in All Five Cateories}
\label{category_five}
\end{table}
%\end{landscape}

%\end{appendix}

\end{document}